\documentclass[twocolumn,trackchanges]{aastex61}

\newcommand{\Msun}{\ensuremath{M_{\odot}}}

\newcommand{\lum}{erg\,s$^{-1}$}

\newcommand\chandra{{\it Chandra}}

\newcommand\kev{{\rm~keV}}

\newcommand {\apgt} {\ {\raise-.5ex\hbox{$\buildrel>\over\sim$}}\ }
\newcommand {\aplt} {\ {\raise-.5ex\hbox{$\buildrel<\over\sim$}}\ } 

\newcommand{\erg}{\rm\; erg}

\received{July 1, 2016}
\revised{September 27, 2016}
\accepted{\today}
\submitjournal{ApJ}
\shorttitle{AGN Feedback in IC~1262}
\shortauthors{M.B. Pandge et al.}
\begin{document}

\title{AGN Feedback in galaxy groups: a detailed study of X-ray features and diffuse radio emission
in IC1262}

\correspondingauthor{Mahadev Pandge}
\email{mbpandge@associates.iucaa.in, mbpandge@gmail.com}

\author[0000-0002-9699-6257]{M. B. Pandge}
\altaffiliation{DST INSPIRE Faculty}
\affil{SERB Young Scientist,Dayanand Science College, Barshi Road Latur, 413 512,  India.}

\author[0000-0001-8985-8596]{S. S. Sonkamble}
\affiliation{National center for Radio Astrophysics (NCRA), Tata Institute of Fundamental Research (TIFR), Pune 411 007, India.}

\author[0000-0001-6282-6025]{Viral Parekh}
\affiliation{Raman Research Institute, C. V. Raman Avenue, Sadashivnagar, Bangalore 560080, India.}

\author[0000-0001-9212-3574]{Pratik Dabhade}
\affiliation{Inter-University center for Astronomy and Astrophysics, Post Bag 4,Ganeshkhind,  Pune-411 007, India.}
\affiliation{Leiden Observatory, Leiden University, Niels Bohrweg 2, 2333 CA, Leiden, Netherlands.}

\author{Avni Parmar}
\affiliation{Raman Research Institute, C. V. Raman Avenue, Sadashivnagar, Bangalore 560080, India.}

\author[0000-0001-6129-8531]{M. K. Patil}
\affiliation{School of Physical Sciences, Swami Ramanand Teerth Marathwada University Nanded, 431 606,  India.}

\author[0000-0002-4864-4046]{Somak Raychaudhury}
\affiliation{Inter-University Centre for Astronomy and Astrophysics, Post Bag 4, Ganeshkhind,  Pune-411 007, India.}
\affiliation{Department of Physics, Presidency University, 86/1 College Street, Kolkata 700073, India.}
\affiliation{School of Physics and Astronomy, University of Birmingham,
  Birmingham B15~2TT, UK.}

\begin{abstract}

This paper reports a systematic search of X-ray cavities, density
jumps and shocks in the inter-galactic environment of the galaxy group
IC~1262 using {\it Chandra}, GMRT and VLA archival observations. The
X-ray imaging analysis reveals a pair of X-ray cavities on the north
and south of the X-ray peak, at projected distances of 6.48\,kpc and
6.30\,kpc respectively.  Total mechanical power contained in both
these cavities is found to be $\sim$12.37$\times 10^{42}$
erg~s$^{-1}$, and compares well with the X-ray luminosity, within the
cooling radius, measured to be $\sim 3.29 \times 10^{42}$
erg~s$^{-1}$, suggesting that the mechanical power injected by the
central AGN efficiently balances the radiative loss. We detect a
previously unknown X-ray cavity at the position of southern radio lobe
in the intra-group medium and find a loop of excess X-ray emission
extending $\sim$100 kpc southwest from the central galaxy.  The X-ray cavity
at the position of southern radio lobe probably represents a first
generation X-ray cavity. Two surface brightness edges are evident to
the west and east$-$north of the center of this group.  The radio
galaxy at the core of the IC~1262 group is a rare low-redshift
ultra-steep radio galaxy, its spectral index being $\alpha\! \sim\!
-1.73$ (including the central AGN) and  $\alpha\! \sim\!
-2.08$ (excluding the central AGN).
We detect a radio phoenix embedded within the
southern radio lobe, for the first time in a poor group, with a spectral index ($\alpha \!\leq\!
-1.92$). The spectral index distribution across the phoenix steepens
with increasing distance from its intensity peak.

\end{abstract}


\section{Introduction} \label{sec:intro}
High resolution X-ray images from the new generation X-ray telescopes,
{\it Chandra} and {\it XMM-Newton}, have provided us with mounting
evidence of various modes of interaction between the hot intergalactic gas,
and the active
galactic nucleus at the core of the clusters (e.g. Perseus), groups
(e.g. NGC 5044) and ellipticals (e.g. Cygnus A). This interaction may
also result in the formation of the substructures or cavities
apparent in the surface brightness distribution of the X-ray emission
\citep{2006ApJ...652..216R,2008ApJ...686..859B,2009ApJ...705..624D,2010MNRAS.404..180D,2011ApJ...735...11O,2012MNRAS.427.3468B,2012A&A...545L...3C}.
Radio observations indicate that many of these cavities are associated
with enhanced level synchrotron emitting cosmic-ray particles of either charge.
As the cosmic rays provide enough pressure without
appreciable increase in the mass density,  these cavities or
bubbles are found to rise buoyantly in the hot gaseous environment. These
cavities can be as small as $\sim$5\,kpc
in diameter, and as large as $\sim$200\,kpc
located in the
central $\sim$20\,kpc of a cluster, or in its outskirts.
They are seen in giant elliptical
galaxies \citep{2002ApJ...567L.115J}, groups
\citep{2006ApJ...648..947M,2008MNRAS.384.1344J,2011ApJ...732...95G,2012ApJ...755..172G}, as well as in galaxy
clusters
\citep{2000ApJ...534L.135M,2004ApJ...607..800B,2006MNRAS.373..959D,2006ApJ...652..216R}. The
mechanical power required for ``inflating" these cavities matches well
with the radiative cooling loss in many of the cluster cores,
suggesting that these bubbles, and perhaps the cosmic rays that fill
them, are involved in the feedback process.
\begin{figure*}
\includegraphics[width=85mm,height=85mm]{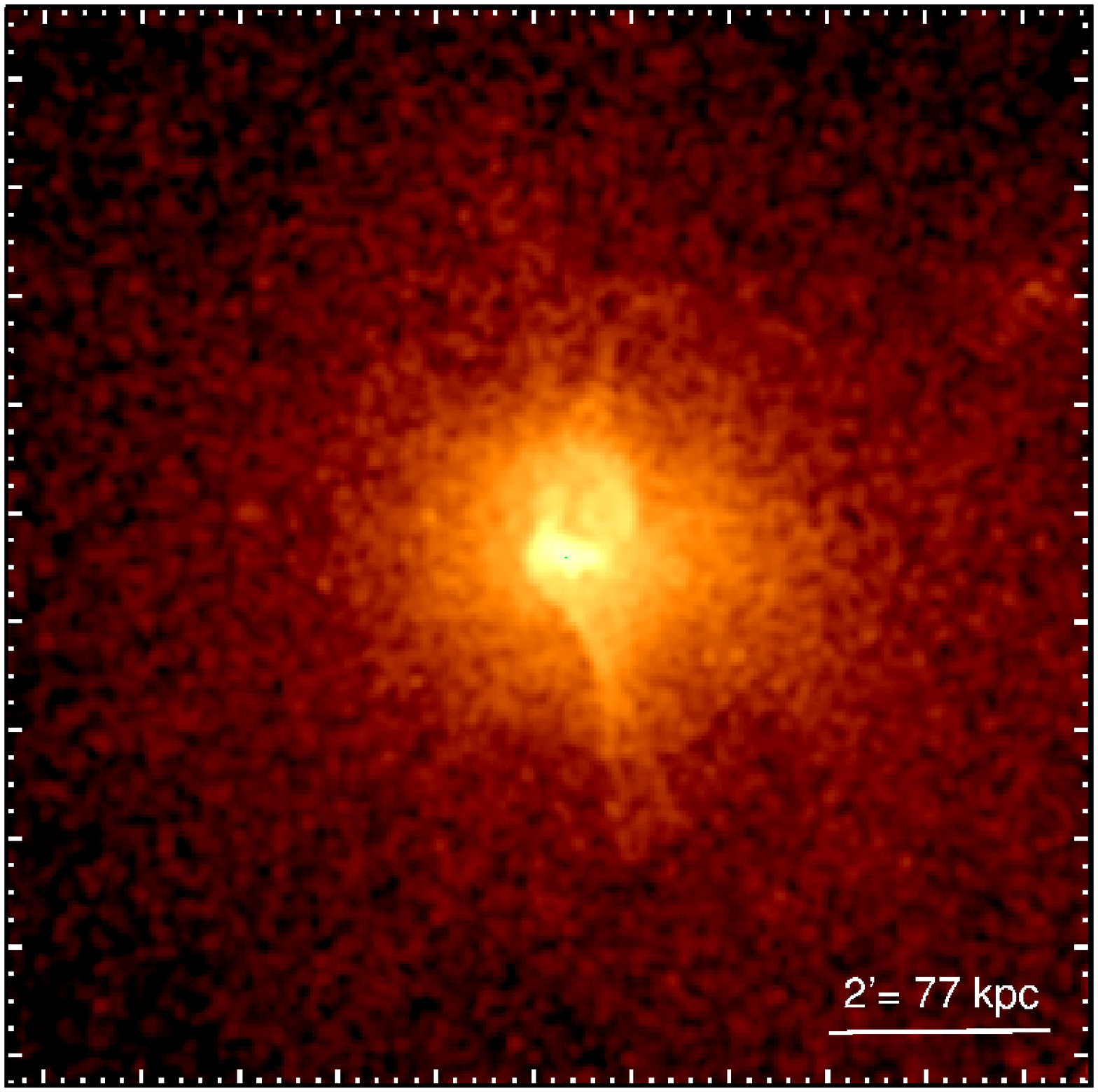}
\includegraphics[width=85mm,height=85mm]{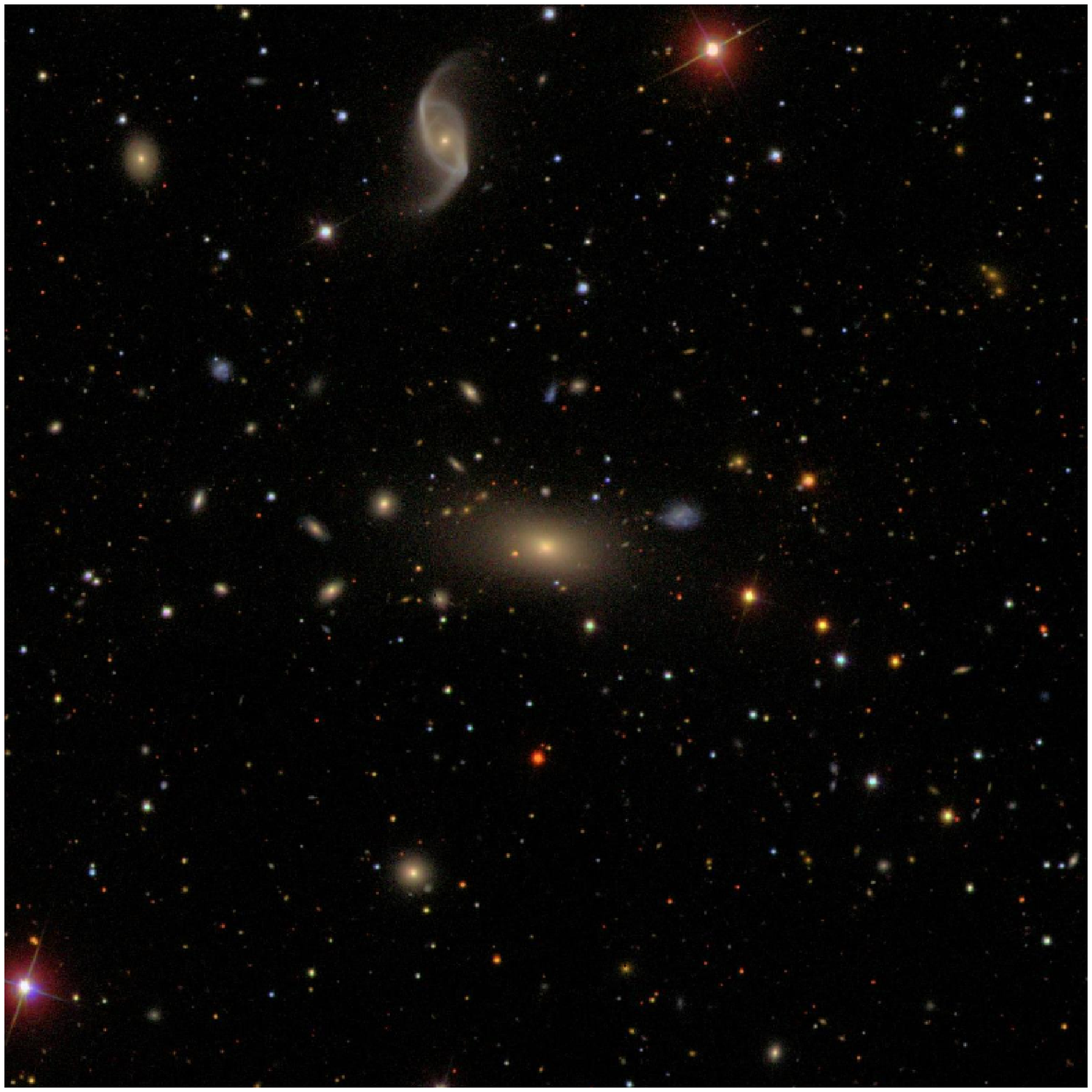}
\caption{{\it Left panel:} The 0.7$-$2.0\,keV background-subtracted,
  exposure-corrected 10$\arcmin\times10\arcmin$ {\it Chandra} X-ray
  image of the IC~1262 group. The X-ray image is smoothed by a 3$\sigma$ Gaussian
  ($\sigma$= 1pixel). {\it Right Panel:} Optical counterpart of IC~1262
  in the R-band, from the Sloan Digital Sky Survey. }
\label{fig1}
\end{figure*}


Most studies of AGN feedback seem to concern massive clusters (e.g.,
the analysis of cluster samples by
\citet{2008ApJ...686..859B,2006MNRAS.373..959D,2009A&A...501..835M},
yet most of the galaxies in the universe are found in smaller systems,
such as poor clusters and groups \citep{2004MNRAS.355..769E}.
Complete volume-limited studies of galaxy groups are rare
\citep[e.g.][]{2017MNRAS.472.1482O}, but these reveal the diversity of
physical phenomena that exist in the intergalactic medium of groups
due to close interactions in the sluggish environment of groups.  A
systematic study of nearby galaxy groups with AGN-ICM interactions is
important to understand the AGN feedback in relatively smaller dark
matter halos. Due to their shallower gravitational potentials, the
AGN outbursts in such systems are believed to produce a larger impact
on the intra-group medium (IGM).

This study is also important because the relationship between the AGNs
and intergalactic hot gas \citep[e.g.][]{2011ApJ...732...95G} and the
atomic and molecular gas \citep[e.g.][]{2018MNRAS.473.5248O} can
significantly influence galaxy evolution in the group
environment. Nearby groups are also useful to probe regions closer to
the central black hole, in the brightest galaxy in the core, in
greater detail, due to proximity and better signal. We have thus
chosen to perform a systematic study of the impact of AGN feedback on
the IGM in the nearby galaxy group IC~1262, which has the eponymous
dominant early-type galaxy, and 31 members within
(20\arcmin$\times$20\arcmin)$^{2}$ \citep{2004AJ....128.1558S}. 

\cite{2000A&A...353..487T} reported a bright arc
  in close proximity of
  the cD galaxy, which may have resulted from the dynamic evolution of the
  central galaxy due to a merger event in this poor group.
  \citep{2003ApJ...595L...1H,2003ApJ...583..706H} studied
  this group using BeppoSAX and {\it Chandra} in the X-ray
  and the NRAO VLA Sky
  Survey and the Westerbork Northern Sky Survey  measurements in the radio,
  to claim the detection of diffuse non$-$thermal emission to
  the South of IC1262, likely associated with a radio mini-halo,
  produced by an earlier merger event.
  These authors also detected a diffuse radio structure of spectral index
  $\sim$1.8, slightly more extended than the
  central cD galaxy.  Further, \cite{2007A&A...463..153T} identified and
  confirmed certain signatures of merging, such as surface brightness
  discontinuities, with disturbed structures in the core and the sharp
  and narrow filamentary structures east of the central galaxy, using
  high resolution X-ray images from {\it Chandra} and XMM$-$Newton. 
  The above studies have in general concluded that high resolution,
  deeper radio observations are needed for IC1262 in order to
  understand the radio properties in more detail. This group is
  located at a relatively low redshift of $z$=0.0326, and its central
  galaxy hosts the radio source 4C+43.46, whose radio power is given
  by log ${\rm~P_{1.4~GHz}} \le 22.56 {\rm~W~Hz^{-1}}$
  \citep{2007A&A...463..153T,2010ApJ...712..883D}. A single X-ray
  cavity was detected by \cite{2010ApJ...712..883D} towards the north
  of the center of the group.  It is interesting to note that the
  central galaxy IC~1262 shows very little star formation ($\sim$
  4.35$\times10^{-2}$\Msun $\rm yr^{-1}$) \citep{2016ApJ...818..182V}.
  In this work, we have tried to assemble available multi-wavelength
  data from different archives such as {\it Chandra}, SDSS, GMRT and
  VLA in order to understand the nature of this group, in particular
  to understand the relation between its merger history and the
  various X-ray and radio features that are detected, and to
  understand the AGN feedback process operateing in this group. In addition,
  we use   high resolution radio images to
  understand the origin of more complex radio structure detected in
  southern radio lobe \citep{2003ApJ...583..706H}.
The structure of the paper is as follows: in Section~2 we present the
data analysis, while the detection of the X-ray cavity is outlined
discussed in Section~3. Section~4 describes the X-ray spectral
analysis. Finally, we present the results from the radio analysis in
Section 5. Throughout this paper we assume $\Lambda$CDM cosmology with
$H_0$ = 70 km\, s$^{-1}$ Mpc$^{-1}$, $\Omega_M$=0.27 \&
$\Omega_{\Lambda}$=0.73, translating to a scale of
0.639\,kpc\,arcsec$^{-1}$ at the redshift $z$=0.032 of IC~1262. All
spectral analysis errors are at 90$\%$ confidence, while all other
errors are at 68$\%$ confidence.

\begin{table*}
\begin{center}
 \caption{{\it Chandra} Observation log}
  \begin{tabular}{llrrrrlrlr}
    \hline
     ObsID &Observing Mode & CCDs on & Starting Date &Total Time (ks)& Clean Time (ks)\\
    \hline
    6949  &VFAINT &0,1,2,3,6  & 2006-04-17 &40&37.54 \\
    7321  &VFAINT &0,1,2,3,6  & 2006-04-19 &40& 35.97\\
    7322  &VFAINT &0,1,2,3,6  & 2006-04-22 &40& 37.47\\
    \hline
    \end{tabular}	
  \label{tab1}
\end{center}
\end{table*}

\begin{figure}
\includegraphics[scale=0.40]{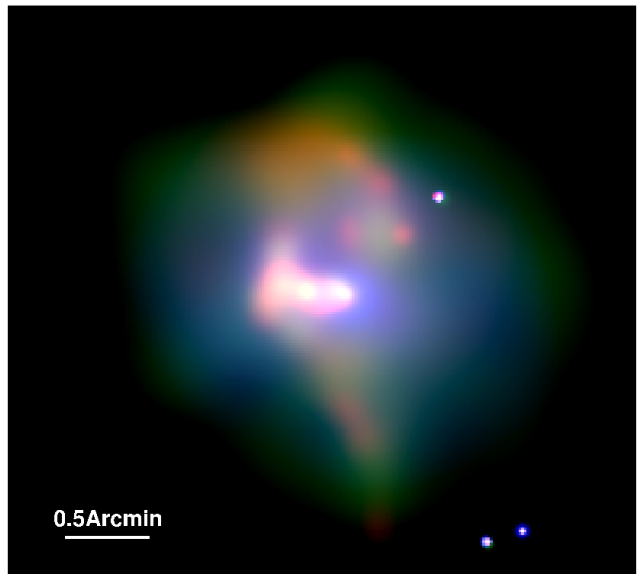}
\caption{Tricolor \chandra\ X-ray image of IC~1262. The soft X-ray image
  ($0.7-1\kev$) is shown in red, the image in the intermediate ($1-2\kev$) band in green
  and that in the hard ($2-8\kev$) band is shown in blue.}
\label{fig2}
\end{figure}

\section{Data Analysis}
\subsection{X-ray observations}
The field of IC~1262 was observed three times in X-rays by the
\chandra~ X-ray Observatory between 17--22 April, 2006, for an
effective exposure of 120\,ks (ObsID and other details in
Table~\ref{tab1}). The observations were reprocessed using
\texttt{{CHANDRA$\textunderscore$REPRO}} task available within CIAO
\footnote{\url{http://cxc.harvard.edu/ciao}} 4.8 and employing the
       latest calibration files CALDB 4.7.2 provided by {\it Chandra}
       X-ray center (CXC). We followed the standard
       \chandra\ data-reduction threads \footnote{\url{http://cxc.harvard.edu/ciao/threads/index.html}} for the
       analysis. Periods of high background flares were identified
       using the {\it lc$\textunderscore$sigma$\textunderscore$clip}
       algorithm, with the threshold set at 3$\sigma$. These periods were removed
       them from the further analysis. The CIAO {\tt
         REPROJECT$\textunderscore$OBS} script was used to reproject
       the event files, and exposure maps in the energy band
       0.7$-$7.0\,keV were extracted using the {\tt
         FLUX$\textunderscore$OBS} script. CIAO scripts {\tt blanksky}
         and {\tt blanksky$\textunderscore$image} were used to identify
       suitable blank sky background fields, corresponding to each of
       the event file, and were used for removing the particle
       background contamination. Point sources were identified using the 
       {\ttfamily WAVDETECT} algorithm within CIAO and were removed
       from the image.  Spectra and corresponding Redistribution
       Matrix Files (RMF), Ancillary Response Files (ARF) were
       generated using the {\tt SPECEXTRACT} task within CIAO 4.8.

\begin{figure*}
\center
\includegraphics[width=80mm,height=80mm]{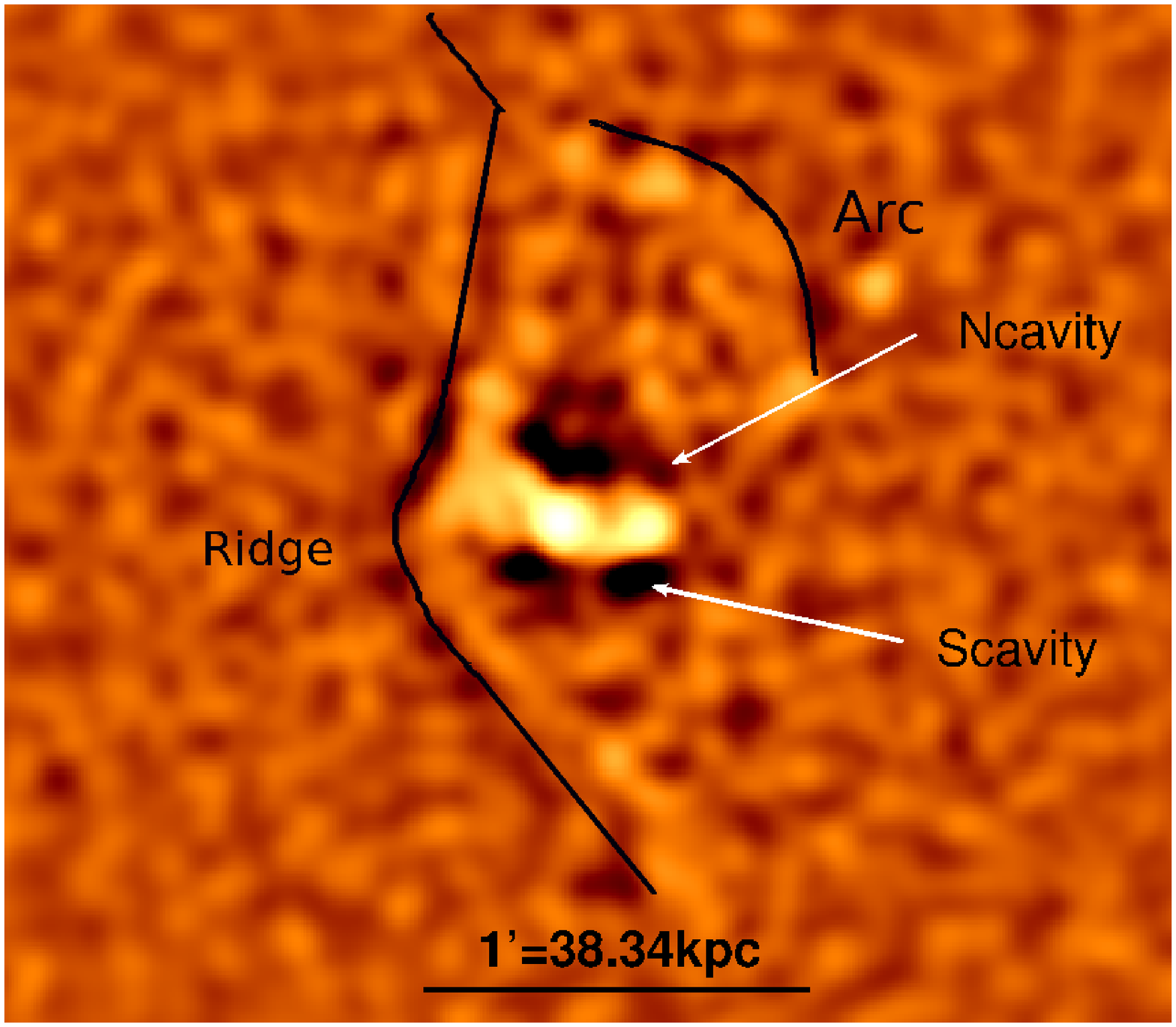}
\includegraphics[width=80mm,height=80mm]{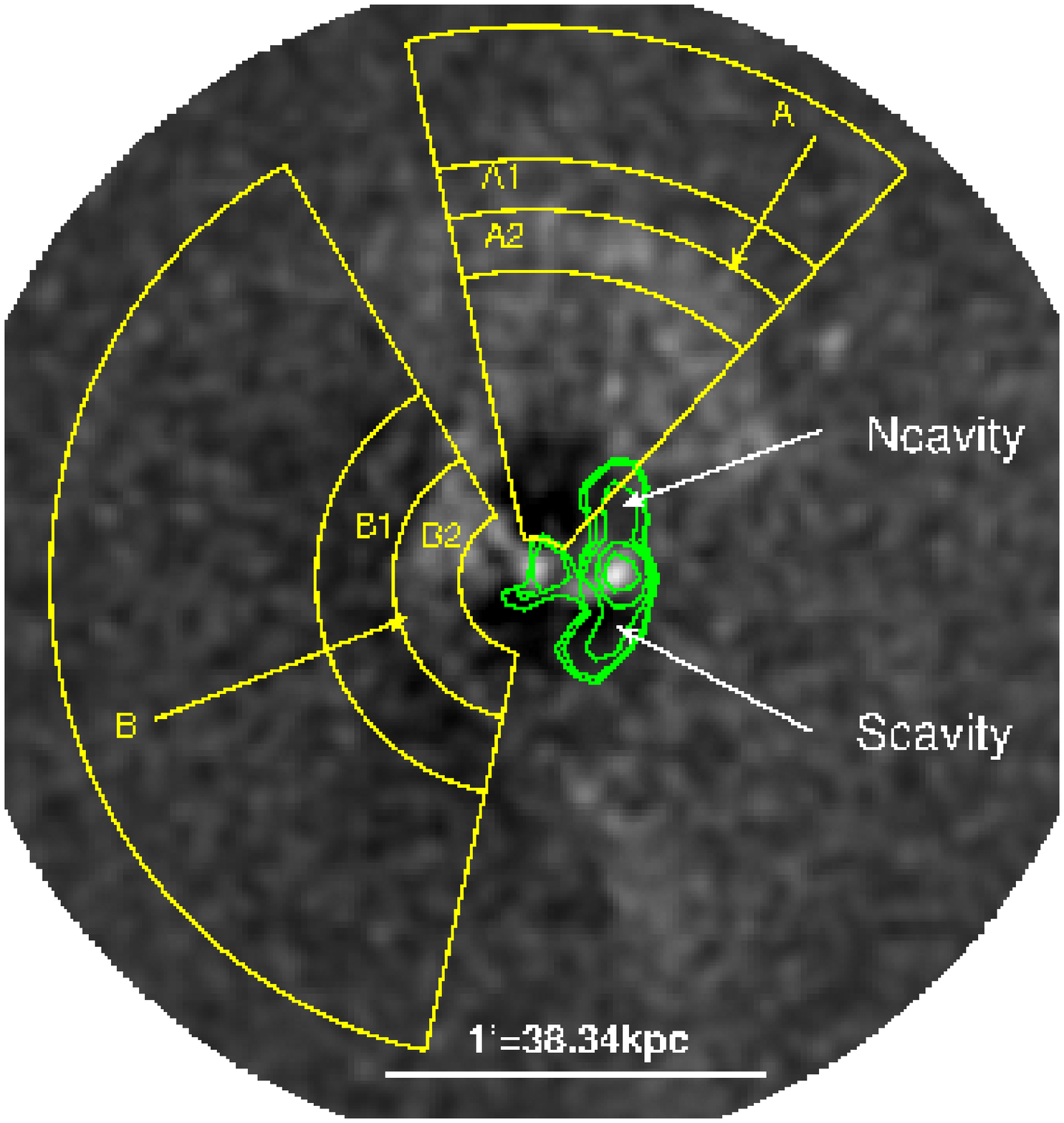}
\caption{ The 0.7-2.0\,keV, unsharp-masked ({\it left panel}) and 2$-$d
  model-subtracted, residual ({\it right panel}) images of
  IC~1262. The unsharp-masked image was derived by subtracting
  a 10$\sigma$ wide Gaussian kernel-smoothed image from that smoothed
  with a 2$\sigma$ wide Gaussian, while the smooth 2$-$d model generated
  from the ellipse fitting was subtracted from the original image to
  derive its residual map. The VLA 1.4~GHz (green) contours are overlaid on
  the residual image. The contours are at (2.5$\sigma$, 3.0$\sigma$,
  3.5$\sigma$) levels where $\sigma$ is 6 $\mu$ Jy $beam^{-1}$ for the VLA
  B-configuration. The annular regions used to extract the surface brightness profiles (A) (30\degr $-$ 110\degr) and (B) (120\degr $-$ 270\degr) are shown by  yellow color.) }
\label{fig3}
\end{figure*}

\subsection{Radio observations}
\label{radio}
\par Archival high frequency VLA and low frequency GMRT data for IC1262 used, whose observational details are summarized in 
Table~\ref{radio_data}.
 \subsubsection{VLA}
 We used VLA archival radio data for the field IC~1262, using L-band continuum (1400 MHz) observations in the B (max baseline 11.1 km) and D (max baseline 1.03 km) configurations.
These observations were analyzed
following the standard routine and using the Common Astronomy Software
Applications (CASA) package of version 4.6.0. The data were inspected
for RFI (radio frequency interference), non-working antennas, bad
baselines, channels and time period. Corrupted data were excised from
the $u$-$v$ dataset. The flux density of each primary or flux
calibrator was set according to \cite{2017ApJS..230....7P}. The
list of both primary and secondary calibrators is given in
Table~\ref{radio_data}.  The same flux calibrator was used for the
bandpass calibration followed by determining the flux density of the
secondary or phase calibrator(s) using the antenna complex gain
solutions. Calibrated visibilities were then used to create the images
by the standard Fourier transform deconvolution. A few rounds of
self-calibration (2 phase + 1 amplitude) were applied to reduce the
effects of residual phase errors in the data and to improve the
quality of the final images. We produced images with the robust `0'
parameter in the 'clean' task of CASA.

 \subsubsection{GMRT}
GMRT archival (Project code {\tt 07MHA01}) P band data of 16 MHz bandwidth centered at 325 MHz frequency and TGSS  150 MHz (TIFR-GMRT SKY SURVEY,\footnote{\url{http://tgssadr.strw.leidenuniv.nl/doku.php}} data \citep{2017A&A...598A..78I}) was used for low frequency study of IC1262. The 325 MHz P band data was processed using SPAM  (Source Peeling and Atmospheric Modeling; \citealt{2014ASInC..13..469I}), an
AIPS-based semi-automated pipeline radio data reduction package that
performs flagging, initial-calibration and imaging with direction
dependent calibration in an iterative way. The complete working of
SPAM can be seen in \cite{2017A&A...598A..78I}. Same package was used for processing the entire TGSS data by \citet{2017A&A...598A..78I}.

%

\begin{table*}
\begin{center}
\scriptsize
 \caption{ IC~1262 radio data.}
 \label{radio_data}
 \begin{tabular}{cccccc}
  \hline
 Telescope & Project code & Frequency (MHz) & Telescope configuration & Date of obs  & Total obs time (min)   \\
  \hline
  VLA & S7601    & 1400  & B  & 27-Jul-2006  & 38 \\
  VLA & S7601    & 1400  & B  & 28-Jul-2006 &  18  \\
  VLA & S7601    & 1400  & D  & 13-May-2007 &  120 \\
  VLA & S7601    & 350   & C  & 11-Jan-2007 &  120  \\
  GMRT & 07MHA01 & 325   & -  & 4-March-2005 & 400  \\
  GMRT & 16{\_}279  & 150 & - & 17-June-2011  & 10   \\
  \hline
  \end{tabular}
   \end{center}
  \end{table*}


\begin{figure*}
\includegraphics[width=85mm,height=85mm]{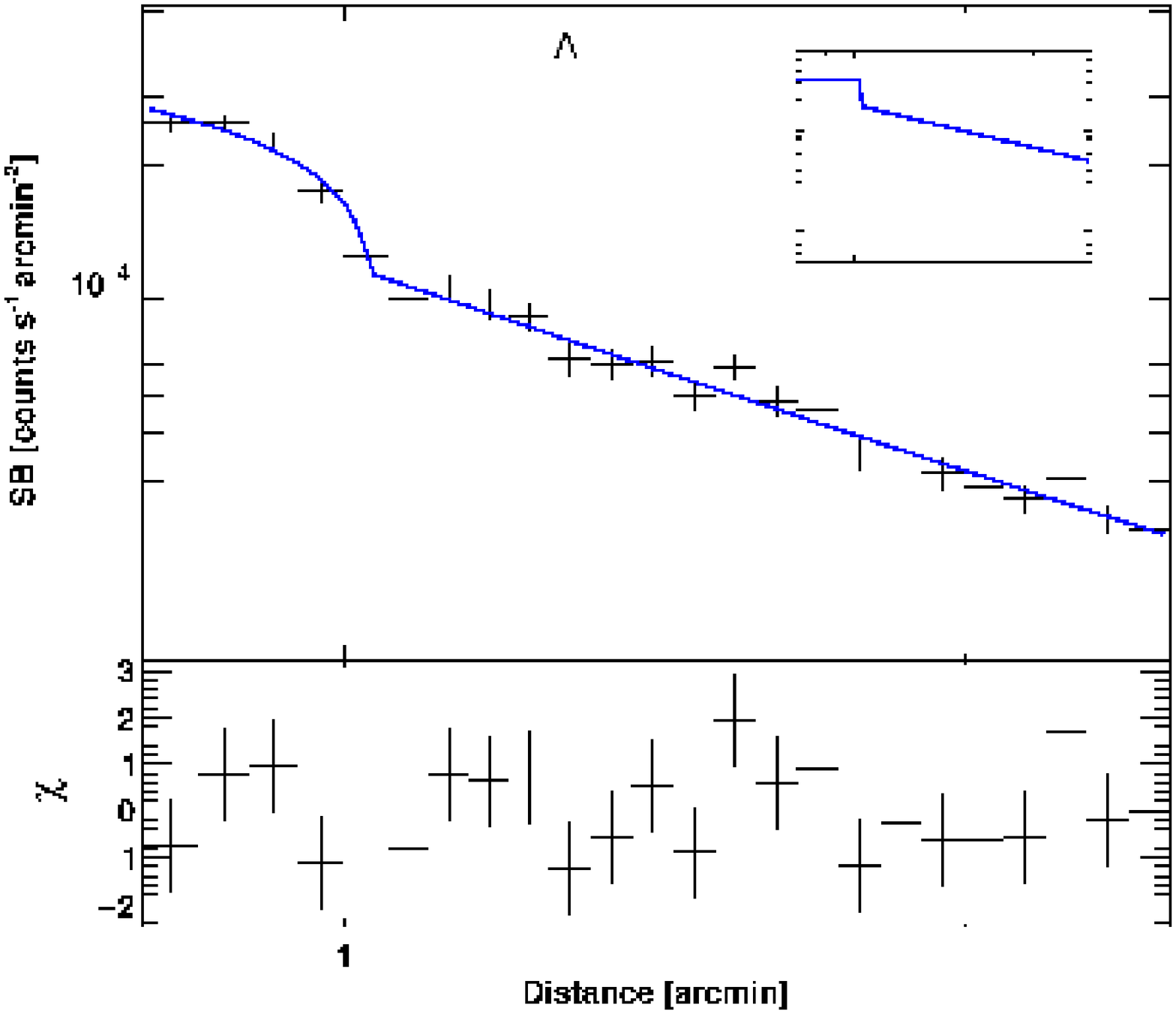}
\includegraphics[width=85mm,height=85mm]{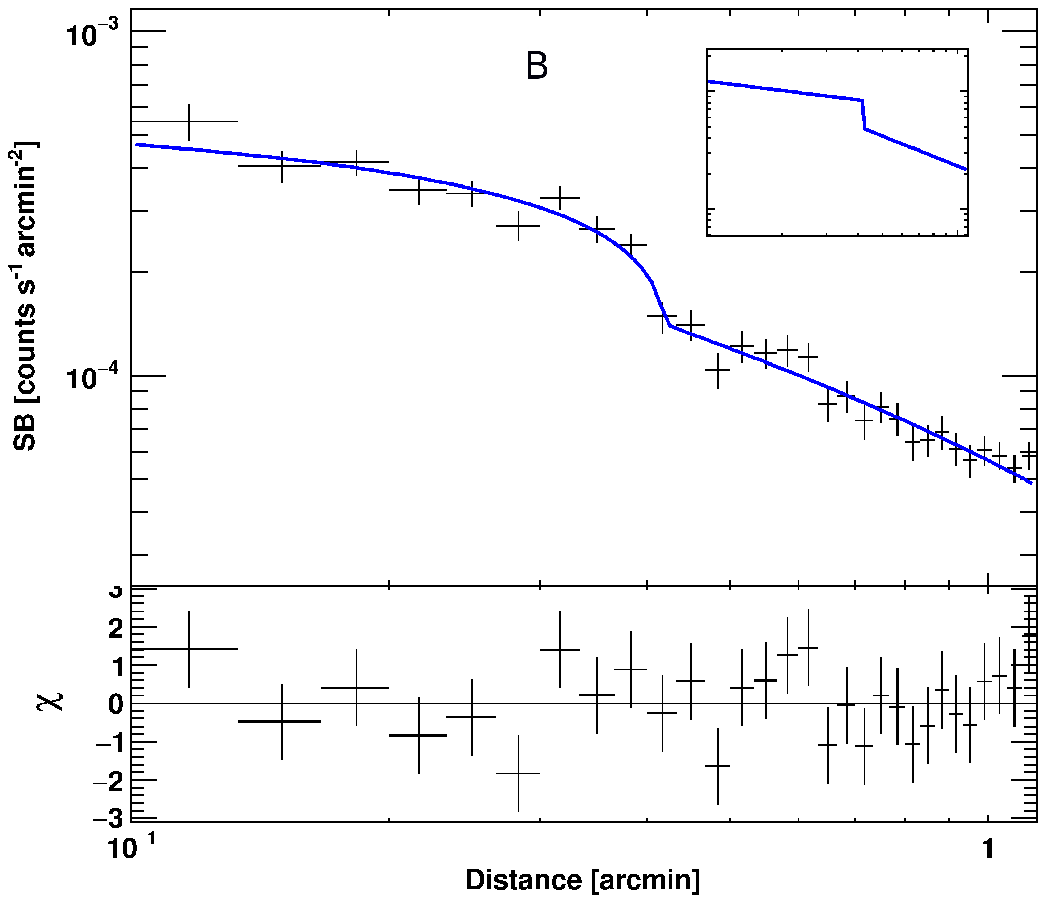}
\caption{Projected radial surface brightness profile, in the energy
  range of 0.7$-$4.0\,keV, extracted from the northwest wedge shaped
  sector with opening angles (30\degr $-$ 110\degr) ({\it left panel})
  and from the east region with angle (120\degr$-$270\degr) ({\it
    right panel}). The surface brightness profiles are extracted from
  the annuli A and B, shown in yellow, in Figure~\ref{fig3}~(A, B). In
  each panel we show the corresponding 3D gas density model, while the
  bottom panels show the residuals from the best-fit surface
  brightness profiles.}
\label{fig4}
\end{figure*}
\section{X-ray Imaging Analysis}
The background subtracted, exposure-corrected, adaptively smoothed
0.7$-$2.0 keV {\it Chandra} image of this group is shown in
Figure~\ref{fig1} (left panel). For comparison, we also show its optical
counterpart R-band image taken from the Sloan Digital Sky Survey
(right panel, SDSS\footnote{\url{http://www.sdss.org/}}) on the
same scale, which confirms the extended nature of the X-ray emission
from this system. This figure reveals interesting structures in the
X-ray image such as surface brightness edges, and a narrow X-ray ridge
in the central region of this group. These structures are consistent with those reported by \cite{2007A&A...463..153T}.

To understand the broad spectral properties of the X-ray map, we
generated a tricolor image in the energy range 0.7$-$8.0\,keV, where,
from the \chandra\ observation, X-ray photons were extracted in three
different energy bands, namely, soft (0.7$-$1.0\,keV), medium
(1$-$2.0\,keV) and hard (2$-$8.0\,keV). These were smoothed, setting
the parameters $sigmin$ = 3 and $sigmax$ = 5, and were combined to
generate the tricolor image for this system (Figure~\ref{fig2}). The
red color in these image represents the soft component of the X-ray
emitting gas, while the medium and hard components are represented in
green and blue respectively. This broadband spectral image highlights
the presence of several substructures in this system.

\subsection{Detection of X-ray Cavities}
\label{cav}
A visual inspection of the 0.7$-$2\,keV \chandra~ image
(Figure~\ref{fig1}) of IC~1262 reveals structures like depressions or
cavities in the surface brightness distribution of the hot ICM, of the
kind reported in \cite{2010ApJ...712..883D}, where the existence of
cavities were discussed, but their position and size were not
measured. Here we identify the cavities and quantify them using
two different techniques.


\subsubsection{Unsharp masked image}
\label{unsharp}
The 0.7$-$2\,keV \chandra~ image of the IC~1262 group, after exposure
correction and subtraction of background, was subjected to a procedure
of unsharp masking, a procedure often used to reveal underlying
structures.  The image was first smoothed with a 2$\sigma$ wide
Gaussian kernel, using the task \textit{aconvolve} within {\tt
  CIAO}. This suppresses the pixel-to-pixel variations of the X-ray
brightness, while preserving small-scale uncorrelated structures.  A
similar image was also generated by smoothing it with 10$\sigma$ wider
Gaussian kernel, which erases small-scale features, while preserving
the overall morphology of the hot gas distribution in this group. The
unsharp masked image was then generated by subtracting the image
smoothed with 10$\sigma$ from that smoothed with 2$\sigma$ wide
Gaussian kernel. The resulting unsharp masked image is shown in
Figure~\ref{fig3} (left panel). A careful inspection of this figure
confirms the presence of X-ray depressions or cavities in the surface
brightness distribution of the hot gas from IC~1262. In addition to
the obvious depressions or cavities, there are some regions
delineating excess emission, as compared to the average surface
brightness.

\subsubsection{$\beta$-model subtracted residual maps}
Further confirmation of the X-ray cavities and excess emission evident
in the unsharp masked image, a 2-D $\beta$-model-subtracted residual
image of IC~1262 was generated. For this, we first generated the 2-dimensional
smooth model by fitting ellipses to the isophotes in the clean
background-subtracted, point-source-removed X-ray image, using the
fitting package {\it Sherpa} available within {\tt CIAO}. The model
parameters i.e., ellipticity, position angle, normalization angle, and
local background, etc. were kept free during the fit.

The best fit 2-D
model was then subtracted from the background-subtracted, exposure-corrected
\chandra~ image of IC~1262 to produce its residual map,
which is shown in Figure~\ref{fig3} (right panel). This
confirms all the major features evident in the unsharp masked
image. The 1.4~GHz (green) contours (from the VLA observation)
are overlaid on the unsharp-masked
image. The contours are at levels 2.5$\sigma$, 3.0$\sigma$, 3.5$\sigma$,
where $\sigma$ is 6 $\mu$ Jy $beam^{-1}$ for the VLA-B
configuration.  

Both these images confirm the presence of a loop-like
structure or an arc (see Figure~\ref{fig3} left panel) directed $\sim$
38$\arcsec$ north-west of IC~1262. In comparable groups
or clusters \citep[e.g.][]{2010ApJ...714..758G,2011ApJ...728..162D}
dark cavities normally are seen
in pairs, on either side of the central excess emission, and are
believed to be the result of the interaction of radio jets from central
dominant system with the surrounding IGM, in the form of buoyantly
rising bubbles. In this system, we detect
two X-ray cavities towards the north (hereafter Ncavity) and south
(hereafter Scavity) from the center: these are highlighted by white
arrows in both the images. The detected X-ray cavities and their
morphological parameters are tabulated in Table~\ref{tab2}. In this
Table, columns 3 $\&$ 4 show the semi-major and semi-minor axes of
the cavities, while column 5 gives their projected distance from the
center of the group.

\begin{table}
\caption{X-ray cavity parameters}
\centering
\begin{tabular}{l c c c c c l}
\hline
Group     	& Cavity     &   a     &   b     & $R$    \\
         	&            &  (kpc)  & (kpc)	 & (kpc)   \\
\hline
IC\,1262  	& Ncavity  &  2.22  &  1.52 	 &6.48		\\
         	& Scavity  &  4.01  &  2.00 	 &6.13		\\
\hline

\end{tabular}
\footnotesize
\begin{flushleft}
\end{flushleft}
\label{tab2}
\end{table}

\begin{table*}
\caption{Best fit parameters of the broken power-law density model.}
\center
\begin{tabular}{lllllllllll}\hline
Regions              &$\alpha$1&     $\alpha$2     & $r_{sh}$     &$n_0$         & C             &$\chi^{2}$/dof \\
                           &                &                            &(arcmin)       &($10^{-4}$)&                &         \\ \hline
A (30\degr - 110\degr)   &$0.18\pm0.11$ &$1.43\pm0.20$ &$1.02\pm0.02$ &$3.20\pm0.40$ &$1.81\pm0.17$ &18.93/18       \\
B (120\degr - 260\degr)&$0.23\pm0.08$ &$1.30\pm0.02$ &$0.41\pm0.03$ &$6.60\pm0.03$ &$1.52\pm0.02$  &48.96/48       \\  \hline
\end{tabular}
\label{tab3}
\end{table*}

\subsection{Surface Brightness Edges} 
In Figure~(\ref{fig1},~\ref{fig2},~\ref{fig3}), we find clear hints of
surface brightness edges (hereafter SBEs) along the east and
north-west directions, at $\sim$ 38$\arcsec$ and 45$\arcsec$
respectively, from the center of IC~1262.  In order to
investigate the origin of these SBEs, we extract surface brightness
profiles of the X-ray emission in the energy range 0.7$-$4.0\,keV, using {\tt
  PROFFIT}-V1.4~\citep{2011A&A...526A..79E}, in the annular regions
(A) (30\degr $-$110\degr) and (B) (120\degr $-$ 270\degr)
5(shown in yellow in
Figure~\ref{fig3}). The extracted profiles along the east and north-west
edges are shown in Figure~\ref{fig4}. These figures indicate that the
apparent sharp
changes in the surface brightness are due the existence of edges along
the respective directions.

The extracted surface brightness profiles across these edges A and B
were fitted with a broken power$-$law density model.  In both cases,
clear density compressions, at levels of over 90$\%$ confidence, are
evident. The broken power$-$law density model is parametrized as:
\begin{equation} n(r)=\left\{\begin{array}{ll} \mathcal{C}n_{0}~(\frac{r}{r_{sh}})^{-\alpha1}, & \mbox{ if }\hspace{2mm}r <r_{\rm sh}\\\\
\mathcal{}n_{0}(\frac{r}{r_{\rm sh}})^{-\alpha2} , & \mbox{if}\hspace{2mm}  r > r_{\rm sh}\end{array}\right.
\end{equation}
where $n$ is the electron number density as a function of the projected
distance, $n_0$ the density normalization,  $C$ the density
compression factor of the shock, $\alpha$1 and $\alpha$2 the power-law
indices, $r$ the radial distance from the center and $r_{sh}$ the
radius corresponding to the putative edge or cold/shock front.
All the parameters of the model were allowed to vary during the
fit.

The best-fit broken power$-$law density model parameters are
summarized in Table~\ref{tab3}.  In order to determine the nature of
the detected edges at A, B (i.e., shocks or cold fronts), we need to
measure temperatures on their either sides. First, we extract the
spectrum from two regions (${\rm A_{1}}$ and ${\rm A_{2}}$) on either
sides of edge A and fit them with a single temperature {\tt APEC}
\citep{2001ApJ...556L..91S} model by keeping redshift fixed at
0.032. The best fit temperature values for ${\rm A_{1}}$ and ${\rm
  A_{2}}$ are 2.39$\pm$0.40\kev~ and 1.68$\pm$0.05\kev,
respectively. The ${\rm A_{2}}$ region contains cool, dense gas and a
sharp boundary. The derived temperatures from regions ${\rm A_{1}}$
and ${\rm A_{2}}$ are inconsistent with each other and indicate that
the edge A is due to the presence of a cold front and is consistent with the cold front reported by \citep{2007A&A...463..153T}.

We also extract the spectra on either side of the edge B and fit them in the same
way, yielding best-fit temperatures of 2.36$\pm$0.23\kev~ and
1.66$\pm$0.04\kev~, respectively. These derived temperature values
seem to indicate that the edge~B is also a cold front and  consistent with the cold front reported by \citep{2007A&A...463..153T}.
\begin{figure}
\includegraphics[scale=0.4]{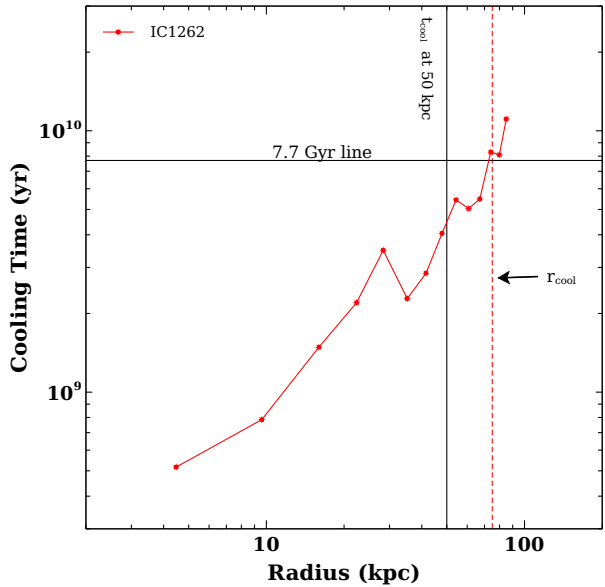}
\caption{Cooling time profile of IC1262. The horizontal
solid line corresponds to the cooling time of 7.7 Gyr. The vertical solid line (black) represents cooling radius at 50kpc, while vertical dashed line (red) represents cooling radius of IC~1262 group where $R_{cool}$=80 kpc at 7.7 Gyr.}
\label{fig7}
\end{figure}
\section{Results and discussion}
\label{c3result}
\subsection{Cooling parameters}
Cooling time is an important parameter to understand thermal evolution
of the hot baryonic matter, in  an individual galaxy, or
in a group or cluster of
galaxies. We derived the {\it cooling time} profile of the X-ray
surface brightness distribution in this group by performing a projected
spectral analysis. The cooling time was then determined for each of
the annular regions, in a projected sense, using the derived values
of the projected electron densities and temperatures. The cooling time
profile was then derived using \cite{1988S&T....76Q.639S}:
\begin{equation}
\label{en:cooling_time}
\rm t_{\rm cool}= 8.5 \times 10^{10} \rm yr \left[ \frac{\rm n_e}{10^{-3} \rm cm^{-3}}\right]^{-1} \left[\frac{\rm T_{\rm gas}}{10^{8}K} \right]^{1/2}.
\end{equation}

where n$_e$ represents the electron density and ${\rm~T_{gas}}$ gas
temperature.  The resultant cooling time profile is shown in
Figure~\ref{fig7}, which reveals that the cooling time of the gas in the
core of this group is much shorter than the Hubble time; in perfect
agreement with those seen in many other groups
\citep{2006PhR...427....1P,2010ApJ...721.1262M,2008ApJ...682..186G,2014MNRAS.438.2341P}. This
strongly supports the argument that heating by the central AGN
occurs at least over a timescale of 10$^8$ yr.

\subsection{X-ray properties within the cooling radius}
To investigate the global properties of the X-ray emitting gas, we
extracted a combined spectrum, from within the cooling radius ($\le
r_{cool}$ 80\,kpc), in the energy range 0.7--8.0\,keV for this
group. The cooling radius is the radius within which gas cools faster
than $7.7\times10^{9}$ yr, the time in which the cluster/group is
believed to relax and establish a cooling flow.

The spectrum was extracted using the {\sc CIAO} tool
{\sc specextract} and grouped to have a minimum of 25 counts per
spectral bin. This was done after removing compact
point-like sources, including the central $\sim2$\arcsec\, region.
The count-weighted response matrices were generated for
each of the extraction.  The extracted spectrum was then exported to
the fitting package {\sc xspec} and fitted with the model ({\tt
  wabs $\times$ apec}). The gas temperature $kT$ and {\sc APEC}
normalization $N$ were allowed to vary during the fit. We repeated the
fitting exercise by freeing and fixing the values of the Galactic hydrogen
column $N_H$ and the gas abundance $Z$. The fit in which $N_H$ and $Z$
were allowed to vary returned in the best-fit results. Using these
best-fit parameters, we derive the 0.7$-$10.0\,keV X-ray luminosity,
from within the cooling radius ($\rm L_{cool}$) for the IC~1262 group, to be
equal to 3.29 $^{+0.2}_{-0.3}\times 10^{42}$\lum.

\subsection{X-ray and Radio morphology} 
\begin{figure}
\center
\includegraphics[scale=0.4]{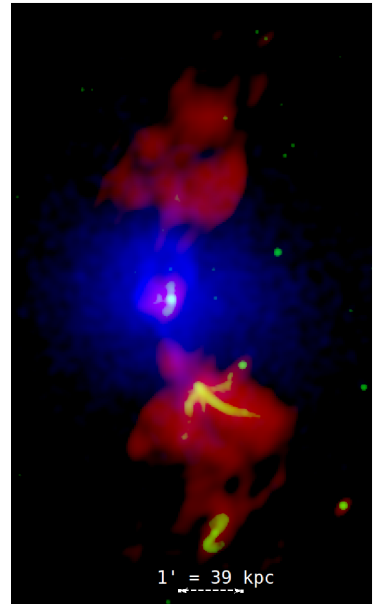}
\center
\caption{The tricolor image of IC~1262, where
  red denotes  325\,MHz radio emission (GMRT), the green
  radio 1.4\,GHz emission at high resolution (JVLA)
  and the blue represents soft X-ray (0.7$-$2.0)\,keV emission from {\it Chandra}.}
\label{fig8}
\end{figure}

 In order to map the diffuse radio emission and its extent in this
 group, we generate a tri-color map, using GMRT radio, VLA radio and
 \chandra~X-ray observations, shown in Figure~\ref{fig8}. In this figure, red
 denotes GMRT 325\,MHz radio emission, green denotes the high
 resolution VLA radio 1.4\,GHz emission map and blue denotes the soft
 X-ray (0.7--2.0)\,keV emission detected by {\it Chandra}. From this
 figure it is evident that radio lobes are more extended than that of
 the X-ray emission. The detailed spectral properties of
 multi-frequency radio emission are discussed in
 Section~\ref{spectral}.

\begin{figure*}
\center
\includegraphics[width=85mm,height=85mm]{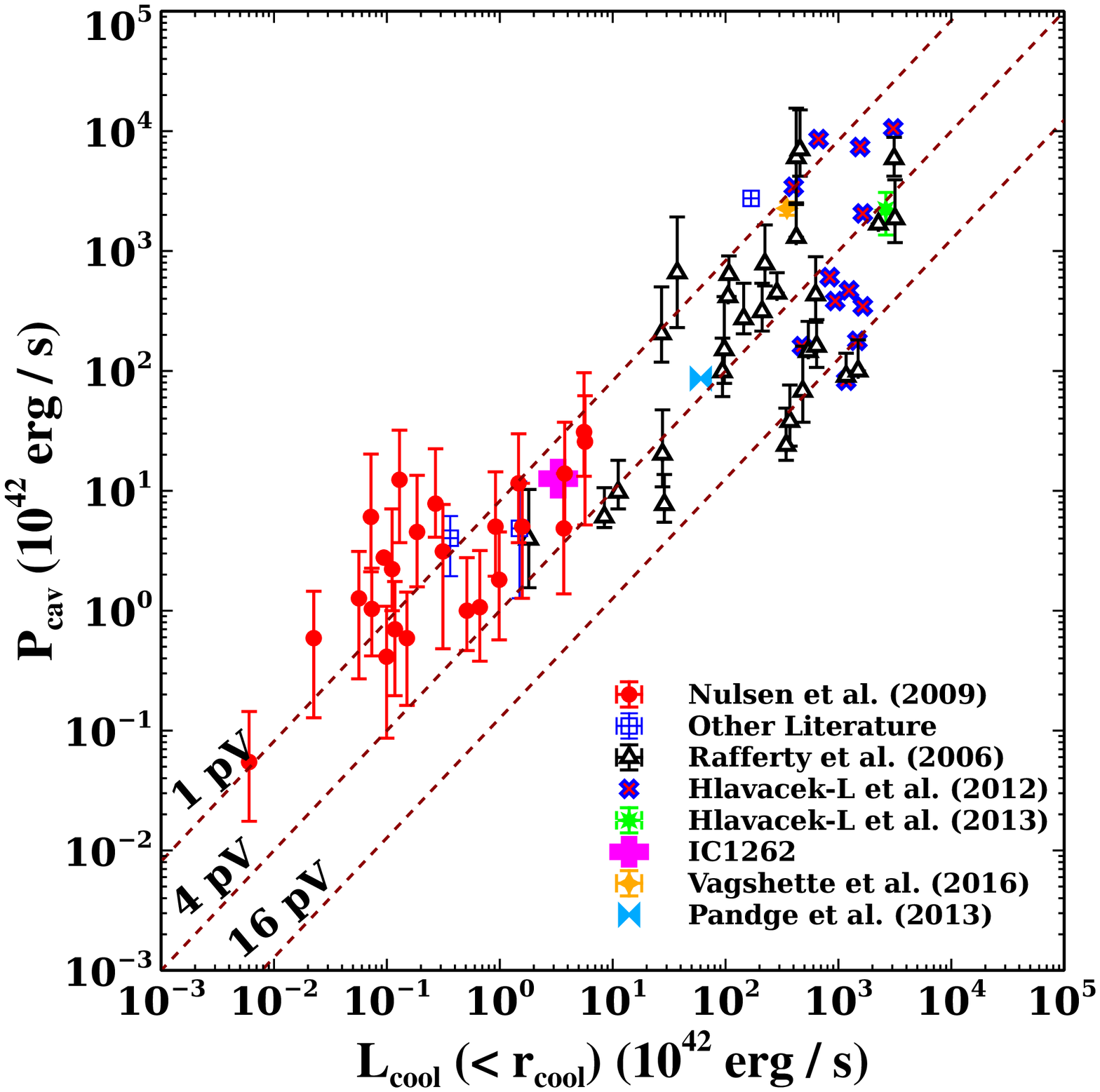}
\includegraphics[width=85mm,height=85mm]{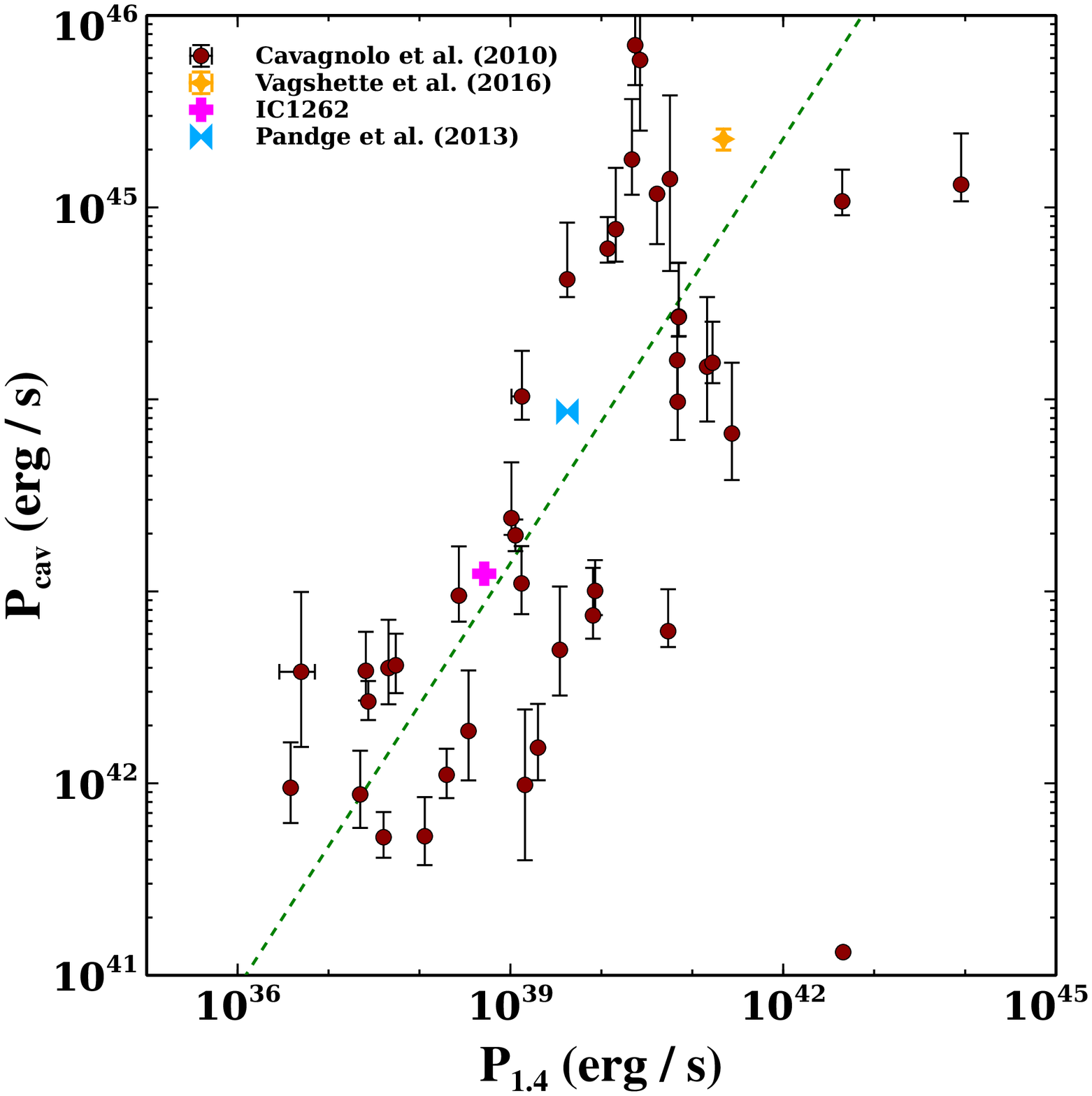}
\caption{{\it left panel:} Cavity power versus X-ray luminosity within the 
  cooling radius of groups and clusters (from \protect\cite{2006ApJ...652..216R}),
  represented
  as filled circles (black). The cavity power and X-ray luminosity for
  IC~1262 is shown by the ``+'' symbol (magenta).  The diagonal lines represent
  samples where $P_{cav}=L_{ICM}$, assuming pV, 4pV, 16pV as the total
  enthalpy of the cavities. {\it Right panel:} Cavity power plotted against radio
  power, with filled circles representing galaxy clusters and groups
  from \protect\cite{2010ApJ...720.1066C}. The magenta ``+'' represents
  the IC~1262 group.}
\label{fig9}
\end{figure*}

\subsection{Cavity Energetics}
 
\begin{table*}
\begin{center}
\caption{Cavity parameters from the X-ray observations}
\begin{tabular}{lccccccccccccc}
\hline
Group    & Cavity       & $E_{\rm bubble}=pV$ & $t_{\rm c_s}$  & $t_{\rm buoy}$  & $t_{\rm refill}$ & $t_{\rm avg}$ & $P_{\rm cav}$ \\          
   (1)      &  (2)       & (3)    &  (4)       &(5)               & (6)     &(7)      &(8)  & \\         
&             & $10^{56}\erg$               & $10^7$ yrs     & $10^7$ yrs      & $10^7$ yrs  & $10^7$ yrs & $10^{42}\erg~s^{-1}$ \\ 
\\
\hline
IC\,1262  & Ncavity     & 58.0 	   & 1.7 	     & 2.4    & 5.2	&3.1	        & 6.0 \\
                 & Scavity     & 50.1	       & 1.2 	    & 2.1		& 4.2	&2.5	        & 6.3 \\
\hline
\end{tabular}
\begin{flushleft}
\end{flushleft}
\end{center}
\label{tab5}
{Column 3: total energy stored in each of the cavity. \\
Column 4, 5 and 6: cavity ages estimated by three different ways (see text). \\
Column 7: average age of cavities. \\
Column 8: power stored in each cavity.}
\end{table*}
\begin{figure*}
\center
\includegraphics[width=180mm,height=120mm]{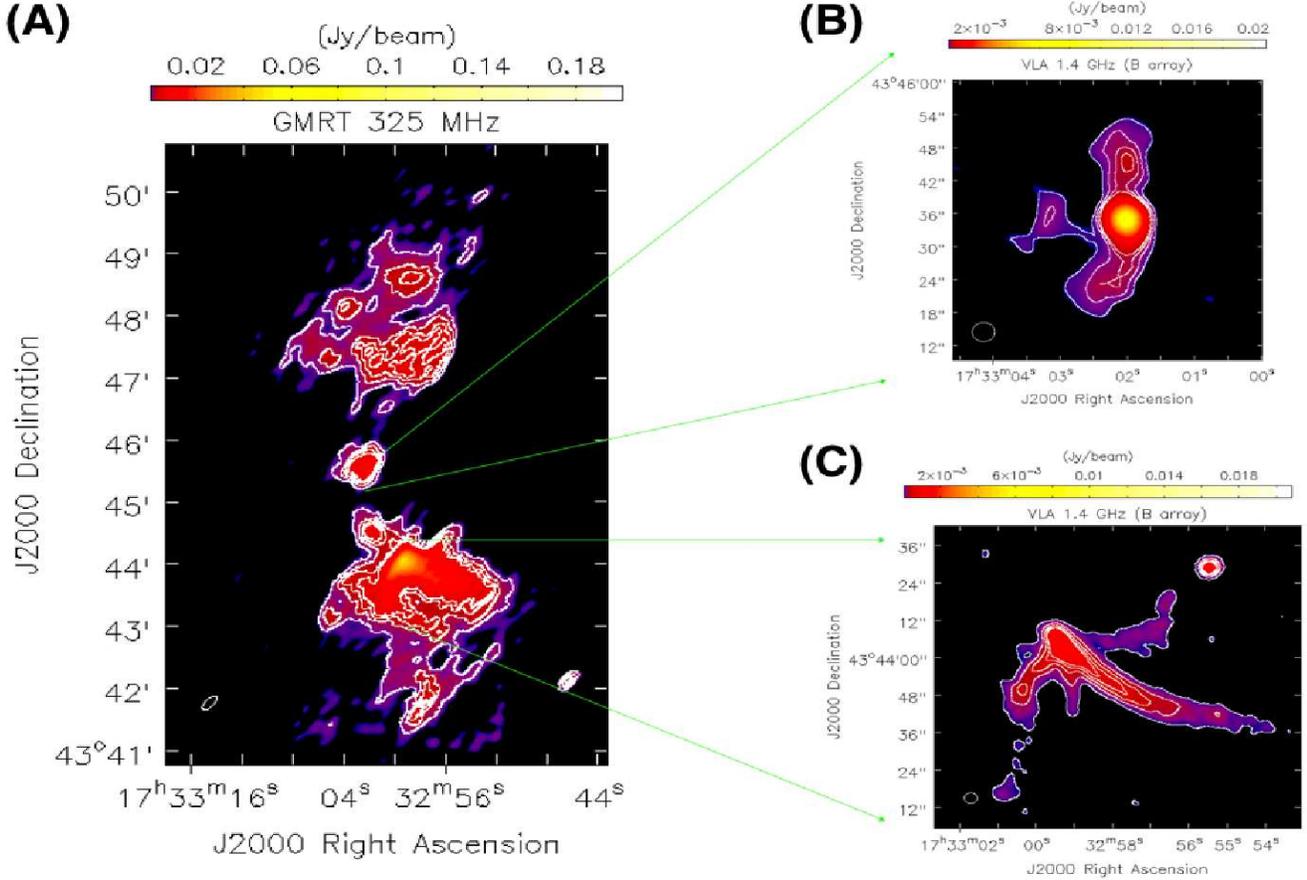}
\caption{color \& contour maps of IC~1262 made, using the task
  \textquotedblleft viewer\textquotedblright~in CASA with relative
  contour levels at [0.2, 0.4, 0.6, 0.8] mJy/beam. Image (A) shows the
  GMRT 325 MHz data with a beam size of
  16.40$\arcsec\times7.39\arcsec$ , where we can see the entire radio
  source at rms of $\sim$ 0.25 mJy/beam. On the right-hand side of the
  image, we see the zoomed-in version of the various components of the
  source. Images (B) and (C) are higher resolution images from JVLA
  (B-array) at 1.4 GHz with a beam size of 3.62$\arcsec\times3.48\arcsec$
  and rms of 0.02 mJy/beam. Image B shows the inner core and jet
  structure. Image C shows the southern lobe of unusual structure.}
\label{fig10}
\end{figure*}
\begin{figure*}
\center
\includegraphics[scale=0.95]{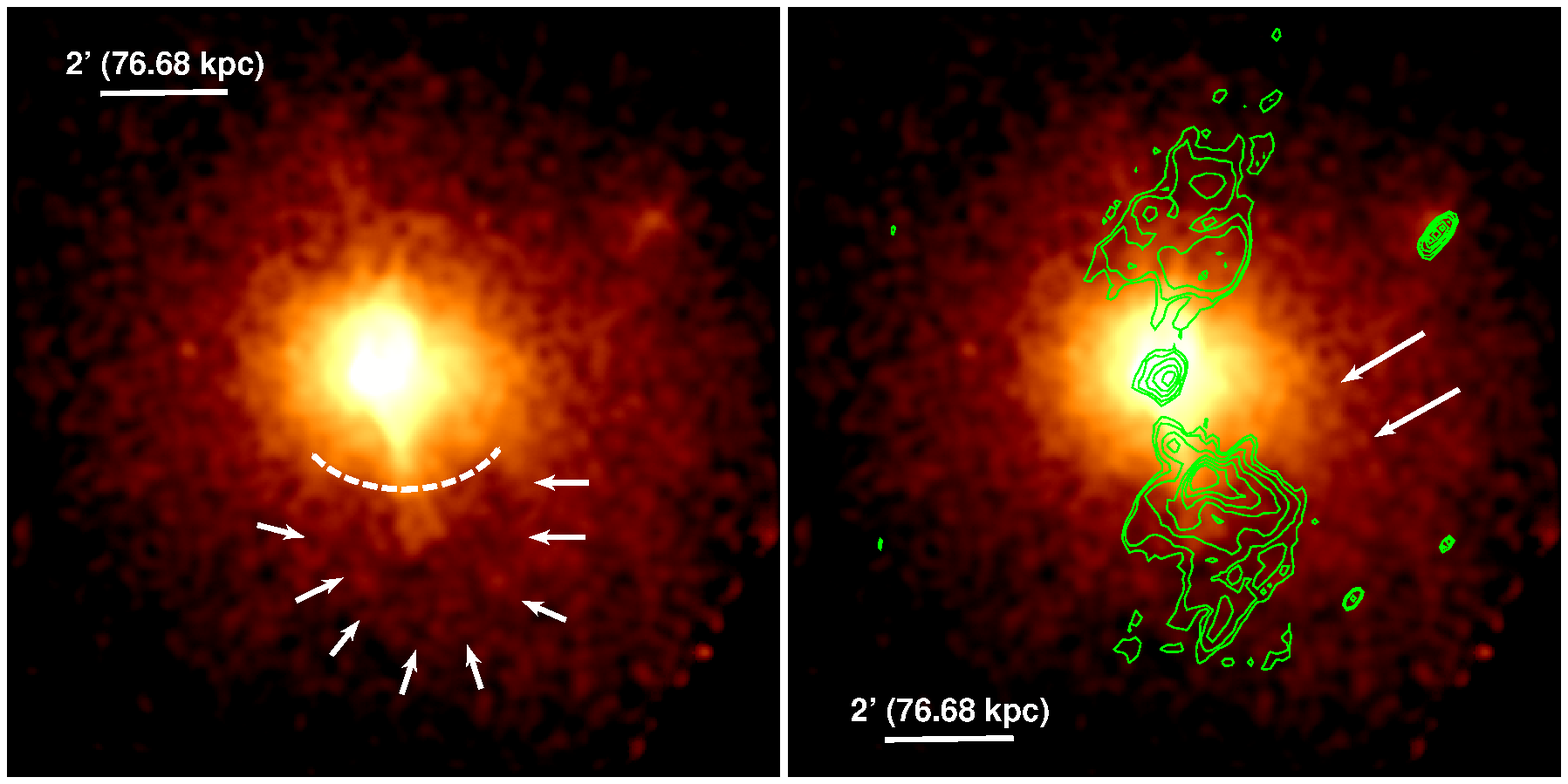}
\caption{{\it left panel:} The {\it Chandra} image of IC1262,
  corrected for exposure and background, in the energy range
  0.7-7.0\,keV. This X-ray image is smoothed by a 6$\sigma$-wide
  Gaussian kernel ($\sigma$= 1 pixel). The location of a first generation
  X-ray depression is shown by white arrows, while the position of the shock is
  shown by a white dotted arc. {\it Right panel:} Same figure as
  left panel, on which GMRT 325~MHz radio contours are overlaid. The
  excess X-ray emission seen southwest of the central source is
  highlighted by white arrows.}
\label{fig11}
\end{figure*}

In the present study we assume the X-ray cavities to be bubbles devoid of
gas, moving outward due to buoyant force at the local ambient
temperature. Volumes of the individual cavities were calculated
assuming that they are symmetric about the plane of the sky, with their
centers lying in the plane perpendicular to the line-of-sight passing
through the central AGN. For this, we assumed the cavities of prolate
ellipsoidal shape, with semi-major axis $a$ and semi-minor axis
$b$. The volume of each  cavity was estimated as $V = 4\pi a b R/3$,
where $R$ is the line-of-sight distance between the nuclear
source and the center of the cavity.
 
Following the method proposed by \cite{2004ApJ...607..800B}, we
estimate the age of the cavities as given in column 7 of
Table~\ref{tab5}. This analysis has enabled us to investigate two
clear cavities with their projected distance from the center of group
equal to $\sim$6.48\,kpc and $\sim$6.13\,kpc.  Generally, the age
derived using the approach of sound crossing provides lower estimates,
while that derived from time required to refill the cavities lead to
estimates on the higher side.
The estimates of the total mechanical power content
of the cavities for the IC~1262 group
are given in column~8 of Table~\ref{tab5}. 
The total mechanical power contained in these two cavities is about
12.37$\times10^{42}{\rm~\erg~s^{-1}}$.

\begin{table*}
\centering
\caption{Radio parameters.}
\begin{tabular}{ ccccccc } 
\hline
  Region & Frequency          &   Flux density & Angular size (deconvolved)   & Linear size &$P_{1.4GHz}$  \\
         &    MHz             &       mJy     &   $''$ $\times$ $''$    &  kpc $\times$ kpc & 10$^{23}$(W Hz$^{-1}$) \\      
\hline
{North Lobe (NL)}   & 150  & 1135$\pm$4.6  &113$\times$90   &68$\times$54\\ 
					& 325   & 341.2$\pm$0.25  & 97$\times$86 &58$\times$22\\ 
					&1400   & 13.04$\pm$0.06  & 97$\times$67 &58$\times$40 &0.3\\ 
\hline
{Central Source (CS)} & 150  & 418.3$\pm$4.5  & 86$\times$37 &51$\times$22 \\ 
					   & 325  & 112.1$\pm$0.25 & 78$\times$24&47$\times$14 \\ 
						&1400 & 16.40$\pm$0.06  &28$\times$18  &17$\times$10& 0.3\\
\hline
{South Lobe (SL)}  & 150 & 2140$\pm$4.6 & 92$\times$75& 55$\times$45\\ 
				   & 325  & 855.1$\pm$0.25 & 76$\times$71 &46$\times$43\\ 
                  &1400  & 50.01$\pm$0.06  &53$\times$44 & 32$\times$27&1.0\\
\hline
\label{tab7}
\end{tabular}
\end{table*}

\subsection{Quenching of the cooling flow}
To cross check whether the apparent AGN feedback can
efficiently work to quench the cooling flow in IC~1262, we compare the
balance between the total AGN output power ($P_{cav}$), derived
from the gas luminosity within the cooling radius
$L_{ICM} (< r_{cool}$). This is shown in the heating versus cooling
diagram (Figure~\ref{fig9} left panel). Here, $P_{cav}$ gives the
measure of the energy injected by the AGN into the hot gas, while
$L_{ICM} (< r_{cool}$) represents the energy lost by the hot gas from
within cooling radius in the form of X-ray emission. This comparison
reveals that the radio source hosted by the central dominant galaxy
IC~1262 is capable enough to deposit sufficient energy into the IGM to
offset the cooling flow.

The equality between the heating and cooling at the heat input rates
of $pV$, $4pV$ and $16pV$ per cavity are shown by the diagonal lines
in this figure. In the same figure we also plot the similar results
derived by other authors
\citep{2006ApJ...652..216R,2011ApJ...735...11O,2013Ap&SS.345..183P,2016MNRAS.461.1885V,2017MNRAS.466.2054V}.
The position of the IC~1262 group in this plot, with respect to the
others from the literature, is shown by a magenta ``+'', and is found
to lie near the 4$pV$ enthalpy line. This in turn confirms that the
total power available within the cavities is sufficient to offset the
cooling flow within the cooling radius.

We also compared the $P_{1.4}$ GHz radio power with total cavity power ($P_{cav}$) (Figure~\ref{fig9}, right panel). In this figure, filled circles represent the sample studied by \cite{2010ApJ...720.1066C}, and IC~1262 is marked by a magenta ``+'' symbol. This plot also indicates that the radio source associated with
IC~1262 is capable of quenching the cooling flow in this group.

\subsection{Association of X-ray cavities with radio emission}
It is well established that the X-ray surface brightness depressions
seen in a majority of the cool-core systems are produced due to the
interaction between the radio lobes, originating from the active
nucleus, and the surrounding ICM, even in galaxy groups
\citep{2008MNRAS.384.1344J,2011ApJ...732...95G,2012MNRAS.421..808P,2013Ap&SS.345..183P,2016MNRAS.461.1885V}.
The radio jets emanating from the cores of the groups or clusters
displaces the surrounding hot gas and leaves depressions or cavities
in their surface brightness. These X-ray cavities are often found to
be filled with radio emission in many of the systems.

We have attempted to examine whether the X-ray cavities in this group
also have a similar association with the radio emission from the
central source. For this we have made use of the multi-frequency radio
data available in the VLA and GMRT archives.  Results from the
analysis of the radio data on this system are shown in
Figure~\ref{fig10}, where Panel A shows the GMRT extended radio emission
map at 325\,MHz, and its contours from the central source, while
Panels B and C are higher resolution images from the VLA (B array) at
1.4 GHz with a beam size of 3.62$\arcsec\times3.48\arcsec$ and rms of
0.02 mJy/beam. Image B shows the inner core and jet structure whereas
image C shows the possible southern lobe that has highly unusual
structure.

The morphology of the radio emission varies greatly over the images
and appears to be quite complicated. Radio emission from IC~1262
appears to be widespread, extending over more than 5\arcmin ($\sim$200
kpc). The radio source in IC~1262 is found to be associated with both
the cavities detected in X-ray image of this group and appears to
coincide with the X-ray center of this group (Figure~\ref{fig3}, right
panel). It implies the presence of two radio sources, one relatively
weaker object associated with the obvious X-ray bright source, while
the other appearing 14.62\arcsec ($\sim$9.7 kpc) to the east of the
former at about.  This radio image shows a clear jet-like structure in
this system (see Figure~\ref{fig10}~B).

We have also detected another X-ray depression at the position of the
southern radio lobe, which could be the first generation X-ray cavity
and is shown by white arrows in Figure~\ref{fig11} (left panel). GMRT
325\,MHz contours (green color) overlaid on this image are shown in
Figure~\ref{fig11} (right panel).  An excess X-ray emission on the west
to the center highlighted by white arrows is also shown in this
figure. A weak shock to the south of the central source at a projected
distance of 200\,kpc has been detected and is shown by a (white)
dotted arc in the Figure~\ref{fig11} (left panel). A weak shock (of Mach
number 1.5) has been reported at this location by
\cite{2009ApJ...693.1142S}. We could not detect a first generation
cavity at the position of the northern lobe. 

\begin{figure*}
\centering
\hspace*{-0.in}
\includegraphics[scale=0.6]{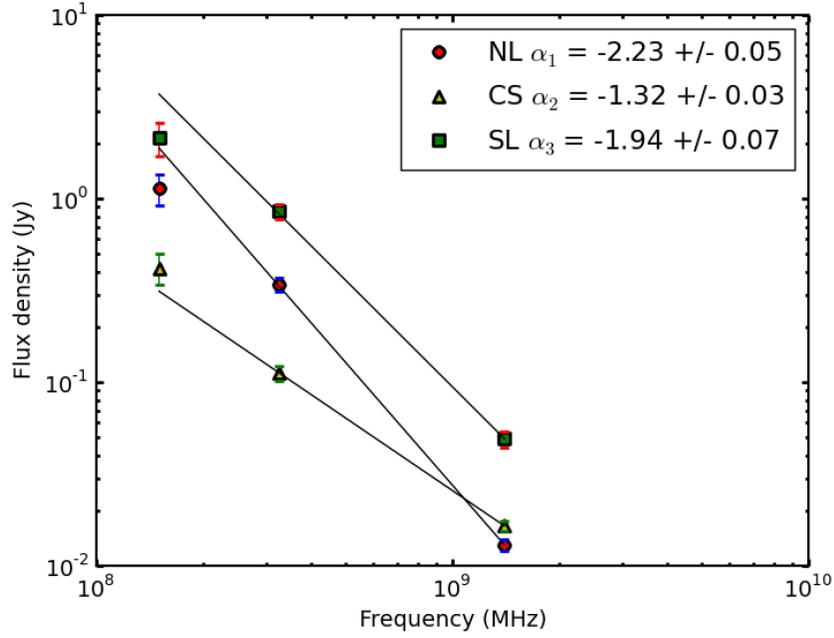}
\caption {The integrated radio spectrum of IC~1262.}
\label{IC1262_spectra}
\end{figure*}
\begin{figure*}
\includegraphics[scale=0.5]{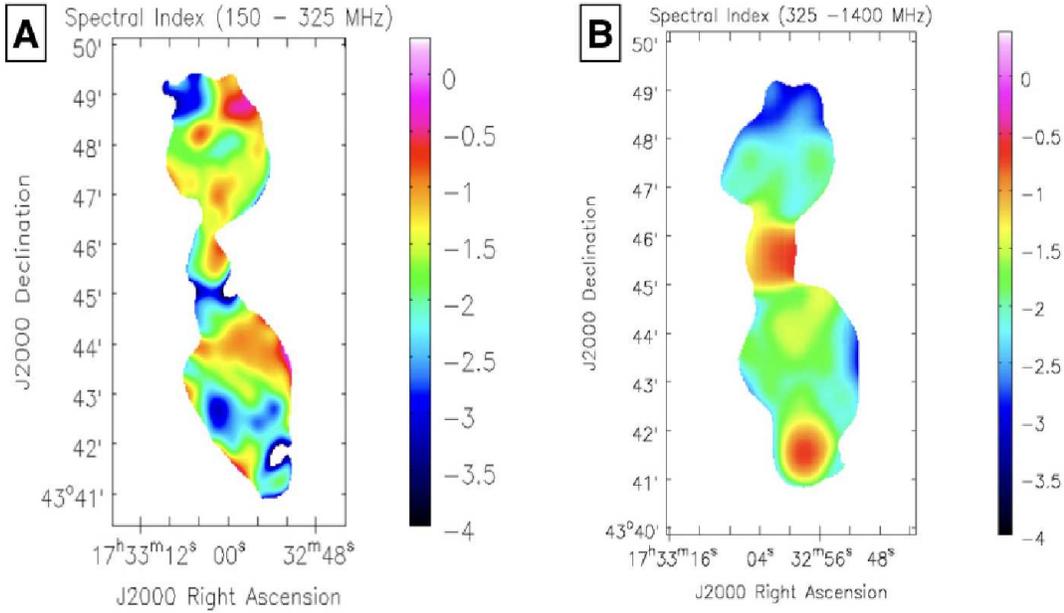}
\caption {{\it A}: Spectral index map between 150\,MHz (TGSS) and 325\,MHz (GMRT) frequencies. {\it B}: Spectral index map between 325 MHz (GMRT) and 1400 MHz (JVLA) frequencies very clearly depicting the steep regions.}
\label{IC1262_si}
\end{figure*}

\begin{figure*}
\includegraphics[width=\textwidth]{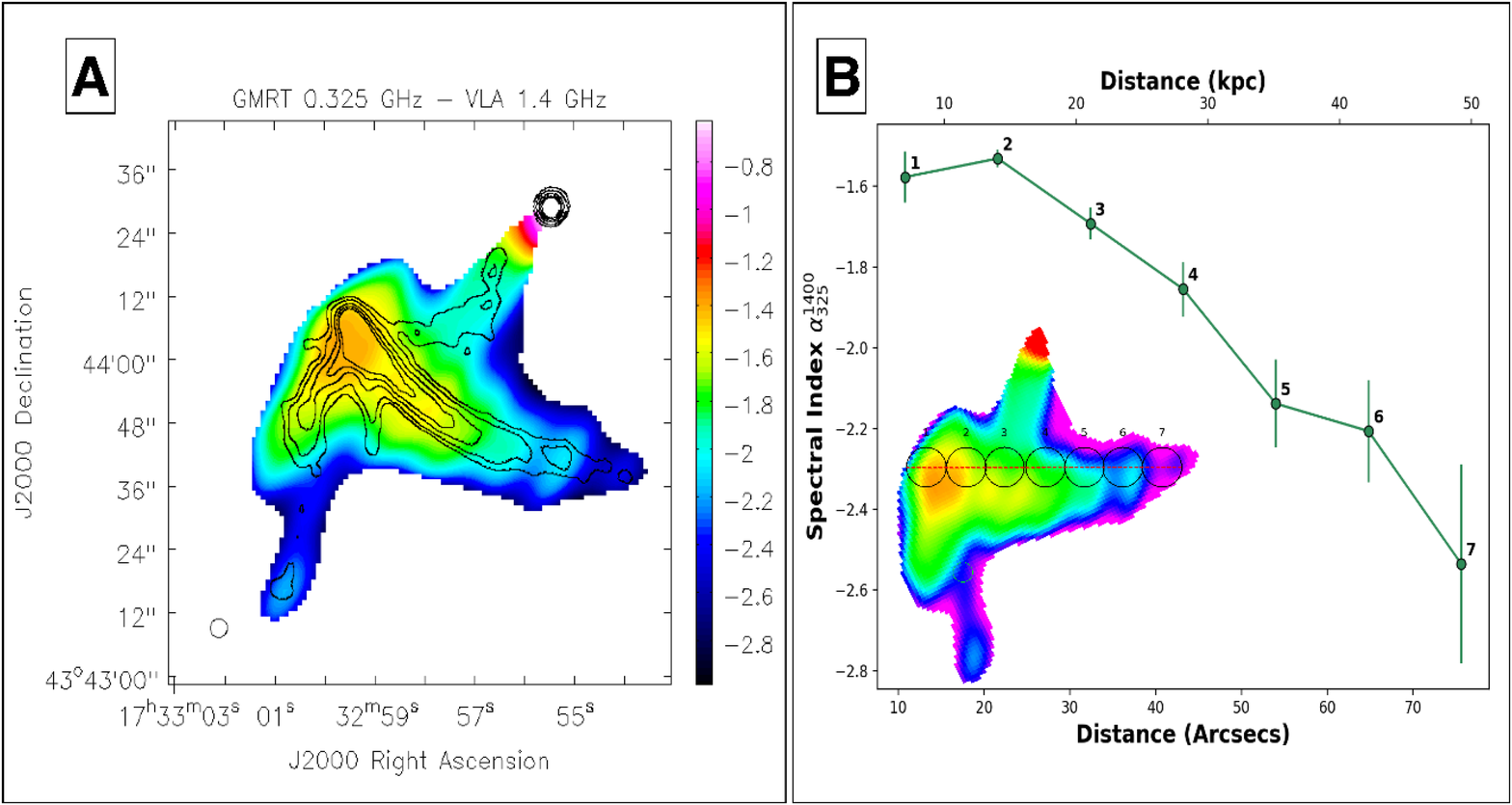}
\caption {{\it A}: The spectral index map of the radio phoenix, measured between
  325\,MHz (GMRT) and 1400\,MHz (VLA) frequencies. {\it B}: Spectral
  index along the radio phoenix, which is embedded in southern radio
  lobe of IC~1262, computed using the regions shown in the inset.}
\label{fig13}
\end{figure*}

\subsection{Spectral index of IC~1262}
\label{spectral}
\subsubsection{Spectral Plot}
\par We measured flux densities for three different regions of IC~1262
- (1) the North Lobe (NL), (2) the Central Source (CS), and (3) the
South Lobe (SL) using observations from
the 150 MHz TIFR-GMRT-Sky Survey (TGSS), VLA P (325~MHz) and
L (1400~MHz) bands.

In the L-band, D-configuration image, we see two separate radio
sources on the south of the central radio source, which are blended in
both the lower resolution P (VLA) and TGSS images. In the flux density
calculation, we separate out this southern source (RA:$17^{h}32^{m}58^{s}$, Dec:$43^{\degr}41^{\arcmin}45^{\arcsec}$) from that in the South lobe 
(RA:$17^{h}32^{m}59^{s}$,
Dec:$43^{\degr}43^{\arcmin}44^{\arcsec}$). We have listed our measurements (deconvolved
sizes and integrated flux densities) in Table~\ref{tab7}.

To calculate the flux densities of these sources, we convolved the
L-band (resolution of 41$''$ $\times$ 37$''$) and TGSS (resolution of
20$''$ $\times$ 20$''$) images to P (VLA) band image of beam size of
57$''$ $\times$ 41$''$. The rms in the L-band, P-band and TGSS images
are 0.15 mJy beam$^{-1}$, 1.5 mJy beam$^{-1}$ and 20 mJy beam$^{-1}$,
respectively. We scaled TGSS, P and L bands flux densities values to
the recent \citep{2017ApJS..230....7P} flux density scale. This new
flux density scale provides accurate measurements over 50 MHz to 50
GHz using current observations with the Jansky VLA. We plotted the
integrated spectra of IC~1262 in Figure~\ref{IC1262_spectra}, where both
the lobes exhibit steep spectrum radio sources (we follow the
convention $S \propto \nu^{\alpha}$, where $\alpha$ is the spectral
index and $S$ is the flux density at frequency $\nu$). We find that
the radio luminosity of the south lobe exceeds that of the north lobe
at all these frequencies.  This is also in agreement with the spectral
index of the north lobe ($\alpha1$ = -2.23) is steeper than the south
lobe ($\alpha3$ = -1.94). Moreover, we find that
average spectral index values for IC1262 radio
galaxy, including and excluding the central AGN,
are 1.73 and 2.08, respectively.

The error estimation in the flux density measurements were carried out
using the following procedure. There are two primary sources of errors
in the flux density measurements: (1) one due to the uncertainties in
the flux densities of the unresolved source(s) used for the
calibration of the data. We assumed this error to be $\sim$ 10$\%$ at
325 and 1400 MHz, while that at 150 MHz to be 20$\%$; (2) since these
radio sources (the north and the south radio lobes) are extended
sources, the errors in their flux density estimations will be the rms
in the image multiplied by the square root of the ratio of the solid
angle of the source to that of the synthesized beam. Since these two
sources of errors are unrelated, we added them in quadrature to
estimate the final error on the flux densities of the extended
sources, using
\begin{equation}
\Delta{S} = [(\sigma_{amp}S)^2 + (\sigma_{rms}\sqrt{n_{beams}})^2)]^{1/2},
\label{cal_err}
\end{equation}
where $S$ is the flux density, $\sigma_{rms}$ the image rms noise,
and $n_{beams}$ the number of beams within the extent of the source.

\subsubsection{Spectral Index Maps}
Three spectral index maps were made, combining data from VLA, GMRT, and
TGSS, as shown in Figure~\ref{IC1262_si}. The maps at various frequencies
were convolved appropriately according to the requirements as
described below:

\begin{enumerate}
\item The higher resolution (16.40\arcsec $\times$ 42.13\arcsec) GMRT
  325 MHz map was convolved to the beam size of the TGSS map (25\arcsec
  $\times$ 25\arcsec). This is shown in Figure~\ref{IC1262_si} (A).
\item The GMRT 325 MHz map was convolved to the beam size of the VLA
  L-band (D-configuration) map (40.96\arcsec $\times$ 37.00\arcsec),
  as shown in Figure~\ref{IC1262_si} (B).
\item The high resolution (3.63\arcsec $\times$ 3.28\arcsec) VLA L-band (B-configuration) map
  was convolved to the resolution of GMRT 325 MHz map, as shown in Figure~\ref{fig13} (A).
\end{enumerate}

After the above procedure, the routine tasks in AIPS (OHGEO and COMB)
were used to make the spectral index maps. Pixels below 3$\sigma
~{\rm rms}$ were blanked in images~A \& B were blanked before computing the
spectral index maps. For Figure~\ref{fig13} (A), a higher cut off of 5$\sigma~{\rm rms}$
was observed for higher precision. In Figure~\ref{fig13} (A), the contours of the VLA L-band (B configuration) are overlaid with its beam size represented in
bottom left corner of the image.  From these figures it is evident that the
average spectral index obtained for the northern lobe is $\sim$ -2.23,
and that of the southern lobe is $\sim$ -1.94 (see Figure~\ref{IC1262_spectra}).
Figures. (\ref{IC1262_spectra} and \ref{IC1262_si}) clearly show that the average spectral index obtained for the northern lobe is steeper than that of the southern lobe. The
average spectral index of both the lobes is extremely steep relative
to that found in typical low redshift radio galaxies, namely $\sim$0.7
\citep{1999AJ....117..677B}.

\subsubsection{Radio Phoenix in the southern radio lobe}

Extended radio sources found in galaxy clusters and groups
are interpreted in a variety of ways, in terms of their
inferred age and origin, from observed parameters such as
extent, shape, flux and spectral index. A taxonomy of such sources
has been summarised in \citet{2004rcfg.proc..335K}, in which  
a {\it radio phoenix} appears as an extended
filamentary steep spectrum (spectral index $\it
-1.5$) radio source. These are found in the cores
of groups and clusters, and are interpreted as an indication
of a recent merger. Accretion shocks in mergers of groups and clusters can be found over
regions as large as hundreds of kpc. These appear extended more
frequently at lower than at GHz frequencies.  Radio phoenixes are
start their life as relics resulting from an earlier merger, having a
population of charged particles that have significantly aged such that
synchrotron radiation from them are too faint to detect at high
frequencies \citep[e.g.][]{2001A&A...366...26E}. The accretion shock due to a
recent merger then rejuvenates this faded relic, by compressing the
plasma and re-accelerating the particles to energies such that they
would again be detectable at radio frequencies. the analogy is thus to
a phoenix rising from its embers. The classic example of a radio phoenix is the cluster Abell 85 \citep{2005ApJ...626..748Y}.
  
  The complex feature detected in southern radio lobe
  (Figure~\ref{fig10}~C) is not associated with any X-ray or optical
  emission, but is found to be embedded within the southern radio
  lobe, suggesting that it is related to the central radio AGN.  The
  radio emission is complex and largely filamentary in morphology,
  which usually indicates interaction with the surrounding ICM, and is
  characteristic of a phoenix. The average value of the radio
  spectral index of this feature within the southern radio lobe
  (Figure~\ref{fig10}~C, obtained by combining fluxes from TGSS, GMRT
  325 MHz, and VLA L-band in the B-configuration) is equal to $\sim$ 
  $-1.92$.  As noted above, this is consistent with that of a
  phoenix, where the aged plasma would have been compressed
  adiabatically by merger shock waves boosting the radio emission
  \citep{2001A&A...366...26E,2004rcfg.proc..335K,
    2011MNRAS.414.1175O,2012A&A...546A.124V}.
    
Several radio phoenixes with these characteristic features have been so far identified, but they have all been discovered in rich clusters with high mass or X-ray temperature. Prominent examples are Abell~85 (radial velocity dispersion $=692$ km~s$^{-1}$) and Abell~1033 (radial velocity dispersion $=677$ km~s$^{-1}$, both values taken from \citealt{2016ApJ...819...63R}). However, IC~1262 is considered to be a poor group, its radial velocity dispersion being about $=300$ km~s$^{-1}$ \citep{1999MNRAS.305..259W}, and thus we find that this is the first case where such an object has been found in a group-scale dark halo, showing that merger shocks can revive relic plasma in group-scale mergers as well.

  We also checked the variation of the spectral index along this
  complex structure, computed using the regions shown in the inset
  (Figure~\ref{fig13}~B). This shows that the spectrum of the diffuse
  emission steepens with increasing distance from the peak of the
  radio emission, which indicates that plasma overall has aged and lost
  its energy via synchrotron or inverse Compton (IC) radiation as it
  has moved away from its source. However, the feature in question,
  appears to have been revived by the passing of a weak merger shock
  with a Mach number ${\cal M}$ $\sim$ 1.2  \citep{2009ApJ...693.1142S}, sufficient for compressing the aged radio plasma.  
  
  To find the possible merger
  signature in the X-ray image described above, we postulate that the
  complex filamentary structure within the southern radio lobe
  originates from an aged radio lobe due from the central radio AGN, which
  has been `revived' by a weak merger shock passing through this
  galaxy group. Thus, we classify this complex feature as a radio
  phoenix \citep{2011MNRAS.414.1175O}. 
\section{Conclusions}
We present results based on the analysis of 120\,ks {\it Chandra}
X-ray, SDSS optical, VLA 350/1400 GHz and GMRT 235/610 MHz radio
observations of the galaxy group IC~1262. The objectives of this study
were to identify and confirm the positions of the X-ray cavities and
surface brightness edges present in the hot intra-group medium (IGM)
of IC~1262. We summarize the important results derived from the
present analysis:

\begin{enumerate}
\item 
In the unsharp masked as well as the 2-d $\beta$-model-subtracted
residual images of the hot intragroup medium, we find two X-ray
cavities (Ncavity and Scavity) and a ridge around the center of the
group IC~1262. \\

\item The X-ray cavities are located at  projected distances of
  $\sim$6.48\,kpc and $\sim$6.13\,kpc from the center of IC~1262. \\

\item Two surface brightness edges are evident to the east and
  north-west of the center of this group and confirmed as cold
  fronts.\\
 
\item The total mechanical power of both X-ray cavities $L_{Cavity}$
  and the X-ray luminosity within the cooling radius $L_{cool}$ indicate
  that the total mechanical power emitted by central radio source is
  sufficient to balance the cooling loss in this group.\\

\item The radio emission from the Jansky VLA 1400\,MHz observation
  appears to coincide
  with the location of detected X-ray cavities.\\

\item The analysis of the X-ray cavity images and the estimated cooling
  time enabled us to calculate the mechanical power for the cavities
  to be  P$_{cavity}$ $\sim$ 12.37$\times$10$^{42}$\lum and
  $L_{cool}$ $\sim$3.29$\times10^{42}$\lum, respectively. The comparison
  of these values implied that the radio jet-mode feedback is
  sufficient to quench the cooling losses occurring in this group
  within the cooling radius.\\

\item From the radio imaging analysis, and the spectral index plot and
  maps, it is evident that the radio sources hosted by IC~1262 have
  a flatter spectrum at the core, while that at the lobes appears
  steeper. \\

\item The radio galaxy belonging to the IC~1262 group is the  low-redshift ultra$-$steep radio galaxy detected with a spectral index $\alpha \sim$ -1.73 and $\alpha \sim$ -2.08 with and without the central AGN, respectively.\\

\item The X-ray depression at the position of southern radio lobe has
  been detected in the present analysis. It is likely that it
  represents a first generation X-ray cavity.\\

\item We detect a radio phoenix embedded within the
southern radio lobe, for the first time in a poor group, having  a spectral index ($\alpha \leq$-1.92). Its spectral index steepens with increasing distance from its peak.

\end{enumerate}

\section*{Acknowledgments}
MBP gratefully acknowledges the support from following funding schemes:  Department of Science and Technology (DST), New Delhi under the SERB Young Scientist Scheme (sanctioned No: SERB/YSS/2015/000534), Department of Science and Technology (DST), New Delhi under the INSPIRE faculty  Scheme (sanctioned No: DST/INSPIRE/04/2015/000108).  MBP wishes to acknowledge with thanks the support received from IUCAA, India in the form of visiting associateship. PD gratefully acknowledge generous support from the Indo-French Center for the Pro- motion of Advanced Research (center Franco-Indien pour la Promotion de la Recherche Avan ́cee) under programme no. 5204-2.  The authors gratefully acknowledge the use of computing and library facilities of the Inter-University center for Astronomy and Astrophysics (IUCAA), Pune, India. This research work has made use of data from the \textit{Chandra} Data  Archive, NASA’s Astrophysics Data System (ADS), NASA/IPAC Extragalactic 
Database (NED), High Energy Astrophysics Science Archive Research Center (HEASARC), and softwares CIAO, \textit{ChIPS}, and \textit{Sherpa} provided by the \textit{Chandra X-ray Center (CXC)}. 

\vspace{5mm}
\facilities{SDSS, CHANDRA (CIAO), GMRT, VLA, CXO}
\software{SPAM \citep{2014ASInC..13..469I},  
          PROFFIT \citep{2011A&A...526A..79E}}

\def\aj{AJ}%
\def\actaa{Acta Astron.}%
\def\araa{ARA\&A}%
\def\apj{ApJ}%
\def\apjl{ApJ}%
\def\apjs{ApJS}%
\def\ao{Appl.~Opt.}%
\def\apss{Ap\&SS}%
\def\aap{A\&A}%
\def\aapr{A\&A~Rev.}%
\def\aaps{A\&AS}%
\def\azh{AZh}%
\def\baas{BAAS}%
\def\bac{Bull. astr. Inst. Czechosl.}%
\def\caa{Chinese Astron. Astrophys.}%
\def\cjaa{Chinese J. Astron. Astrophys.}%
\def\icarus{Icarus}%
\def\jcap{J. Cosmology Astropart. Phys.}%
\def\jrasc{JRASC}%
\def\mnras{MNRAS}%
\def\memras{MmRAS}%
\def\na{New A}%
\def\nar{New A Rev.}%
\def\pasa{PASA}%
\def\pra{Phys.~Rev.~A}%
\def\prb{Phys.~Rev.~B}
\def\prc{Phys.~Rev.~C}%
\def\prd{Phys.~Rev.~D}%
\def\pre{Phys.~Rev.~E}%
\def\prl{Phys.~Rev.~Lett.}%
\def\pasp{PASP}%
\def\pasj{PASJ}%
\def\qjras{QJRAS}%
\def\rmxaa{Rev. Mexicana Astron. Astrofis.}%
\def\skytel{S\&T}%
\def\solphys{Sol.~Phys.}%
\def\sovast{Soviet~Ast.}%
\def\ssr{Space~Sci.~Rev.}%
\def\zap{ZAp}%
\def\nat{Nature}%
\def\iaucirc{IAU~Circ.}%
\def\aplett{Astrophys.~Lett.}%
\def\apspr{Astrophys.~Space~Phys.~Res.}%
\def\bain{Bull.~Astron.~Inst.~Netherlands}%
\def\fcp{Fund.~Cosmic~Phys.}%
\def\gca{Geochim.~Cosmochim.~Acta}%
\def\grl{Geophys.~Res.~Lett.}%
\def\jcp{J.~Chem.~Phys.}%
\def\jgr{J.~Geophys.~Res.}%
\def\jqsrt{J.~Quant.~Spec.~Radiat.~Transf.}%
\def\memsai{Mem.~Soc.~Astron.~Italiana}%
\def\nphysa{Nucl.~Phys.~A}%
\def\physrep{Phys.~Rep.}%
\def\physscr{Phys.~Scr}%
\def\planss{Planet.~Space~Sci.}%
\def\procspie{Proc.~SPIE}%
\let\astap=\aap
\let\apjlett=\apjl
\let\apjsupp=\apjs
\let\applopt=\ao
\bibliographystyle{aasjournal.bst}
\bibliography{mybib.bib}
\end{document}